\documentclass[reprint,amsmath,amssymb,aps,showkeys]{revtex4-2}

\usepackage{slashed}
\usepackage{amsmath}

\usepackage{graphicx}
\usepackage{caption}
\usepackage{subcaption}
\usepackage[justification=raggedright,format=hang]{caption}
\usepackage{dcolumn}
\usepackage{bm}
\usepackage{hyperref}
\usepackage[mathlines]{lineno}
\usepackage{xcolor}
\usepackage[makeroom]{cancel}
\usepackage{upgreek}

\newcommand\RR{\mathbb{R}}

\newcommand\id{\textit{id}}

\newcommand\vphi{\varphi}

\newcommand\rarrow{\rightarrow}

\renewcommand\b{\bar }

\renewcommand\-{^{-1}}

\renewcommand\id{\text{id}}

\DeclareMathOperator{\Diff}{Diff}

\begin{document}


\title{Raising galaxy rotation curves via dressing} 

\author{J. François}
\email{jordan.francois@uni-graz.at}
\affiliation{University of Graz (Uni Graz), 
Heinrichstraße 26/5, 8010 Graz, Austria, and \\
Masaryk University (MUNI), 
Kotlářská 267/2, Veveří, Brno, Czech Republic, and \\
Mons University (UMONS), 
20 Place du Parc, 7000 Mons, Belgium. 
}

\author{L. Ravera}
\email{lucrezia.ravera@polito.it}
\affiliation{Politecnico di Torino (PoliTo),
C.so Duca degli Abruzzi 24, 10129 Torino, Italy, and \\
Istituto Nazionale di Fisica Nucleare (INFN), Section of Torino,
Via P. Giuria 1, 10125 Torino, Italy, and \\
Grupo de Investigación en Física Teórica,
Universidad Cat\'{o}lica De La Sant\'{i}sima Concepci\'{o}n, Chile.
}


\begin{abstract}

We present a manifestly diffeomorphism-invariant simple model of galaxy dynamics obtained by applying the Dressing Field Method (DFM) to a general-relativistic system comprising the metric and four scalar fields, phenomenologically representing the four-velocity of a cosmological fluid or dust field. 
The DFM, a systematic tool for extracting the gauge-invariant content in general-relativistic theories, provides a physical coordinatization that yields corrective terms to the rotational velocity profile. 
These corrections produce galaxy rotation curves that combine a Keplerian and a constant velocity terms, effectively emulating a Dark Matter contribution.
We compare DFM-derived rotation curves to observed data for spiral galaxies, from the Spitzer Photometry and Accurate Rotation Curves (SPARC) database, showing that the DFM allows to fit them well.

\end{abstract}


\maketitle


\section{Introduction}
\label{Introduction}

Spiral galaxies rotate around their vertical axes and, by measuring the Doppler shift of atomic lines, one can
determine the circular velocity of stars and other tracers as a function of their distance from the galactic center, obtaining a rotation curve. This produces galaxy rotation curves, which thus describe the orbital speeds of stars and gas in galaxies as a function of their distance from the galactic center.
Typically, Newtonian gravity predicts that these speeds should decrease with increasing distance, with a Keplerian decline. However, observations reveal that rotation curves remain flat -- or even rise -- at large radii, suggesting the presence of ``unseen" mass, thus associated with Dark Matter (DM), which provides additional gravitational influence. 
This discrepancy between theoretical predictions and observed velocities, first noted by astronomers like V. Rubin in the 1970s, supports the hypothesis that DM halos surround galaxies, contributing significantly to their gravitational potential. See, e.g., \cite{Rubin:1978kmz} and the comprehensive reviews \cite{Sofue:2000jx,Bertone:2004pz,Taoso:2007qk,Strigari:2012acq,Cirelli:2024ssz}. The latter provides also a classified collection of literature on various topics related to DM, galaxy rotation curves, and alternative theoretical scenarios (such as, famously, the one of MOND).

In this letter, we establish that a manifestly invariant formulation, obtained through the application of the Dressing Field Method (DFM) \cite{Francois2023-a,JTF-Ravera2024gRGFT,JTF-Ravera2025bdyDFM,JTF-Ravera2025MAG} to a system whose field content is given by the metric and four scalar fields -- representing, e.g., the phenomenological four-velocity of a cosmological fluid or dust field -- yields ``raised" galaxy rotation curves, effectively emulating DM contributions. 

The DFM is a systematic mathematical tool to exhibit the gauge-invariant, relational \cite{JTF-Ravera2024c} content of general-relativistic (gauge field) theories, whereby physical field-theoretical degrees of freedom (d.o.f.) co-define each other and coordinatize the \emph{physical spacetime}.

In this paper, we apply the DFM to galaxy rotation curves, exploring whether a gauge-invariant reformulation of gravitational and matter d.o.f. can naturally produce flat rotation curves, reproducing the effects traditionally attributed to DM.
Our approach thus tests an alternative theoretical framework for galactic dynamics. 
In this cosmological application, the DFM generates a physical scalar coordinatization, which, under specific conditions for the perturbative expansion of the scalar fields, yields a constant corrective term alongside the Keplerian contribution to the rotational velocity squared, therefore raising galaxy rotation curves.

The remainder of this paper is structured as follows: 
In Section \ref{Dressing for diffeomorphisms}, we briefly review the basics of the DFM in the case of diffeomorphisms in general-relativistic physics, following \cite{JTF-Ravera2025bdyDFM}. 
In Section \ref{Physical coordinatization corrections to rotational velocity}, we apply the DFM to the four-dimensional general-relativistic case in which the field content is given by the metric field and matter, described phenomenologically as a fluid supplying scalar fields $\vphi$. The latter, in fact, provide the dressing field to obtain the manifestly
diffeomorphism-invariant formulation, acting as a reference physical system.
We consider the (bare) Schwarzschild metric to provide the baseline geometry for a central mass $M$. 
The scalar fields -- more precisely, the fields appearing in the perturbative expansion of the latter -- contribute to the effective mass $M_{\text{eff}}$. 
These corrections are external to the original Schwarzschild solution -- in the sense that, in principle, they arise from solving the (dressed) Einstein equations with a dust field contribution -- but are incorporated into the dressed metric to describe the \emph{relational}, dressed gravitational field. 
We show that the scalar profile introduced allows for corrective terms that raise galaxy rotation curves, i.e. provide a constant correction to the squared rotational velocity. 
In Section \ref{Scalar profile and four-velocity of dust field}, we discuss the interpretation of the scalar profile in terms of the four-velocity of a dust field. 
In Section \ref{Phenomenological comparison of DFM to observed rotation curves}, we compare DFM-derived rotation curves to observed data for the spiral galaxies NGC 3198 and NGC 2403, using rotation curve data extracted from the Spitzer Photometry $\&$ Accurate Rotation Curves (SPARC) database \cite{Lelli:2016zqa}. 
In the Supplemental Material \ref{Supplemental Material: Further galaxy plots and fits of DFM velocity profiles}, we also provide further fits of DFM velocity profiles for other high-resolution galaxies.
This phenomenological analysis shows that the DFM allows to fit rotation curves, particularly well at large radii.
Section \ref{Conclusions} is devoted to final remarks and future developments.
While this paper focuses on galaxy rotation curves, the DFM framework is extremely versatile, and can be extended to other cosmological phenomena, such as perturbation theory and large-scale structure formation, which we plan to explore in future work.

\section{Dressing for diffeomorphisms}\label{Dressing for diffeomorphisms}

Via the DFM one produces gauge-invariant variables out of the field space $\Phi=\{\upphi\}$ of a general-relativistic (gauge field) theory. 

Let us briefly remind the dressing procedure implemented in the DFM in the case of $\Diff(M)$, being $M$ the manifold on which the field theory is defined. 
Here we provide a field-theoretical presentation, while we refer the reader to \cite{Francois2023-a,JTF-Ravera2024gRGFT} for a formulation in terms of differential bundle geometry.

\smallskip

Consider a general-relativistic theory with field content given by, e.g., $\upphi=\{A, \vphi, g\}$, where $A$ is a 1-form gauge potential, $\vphi$ represents the matter fields, and $g$ is a metric field on $M$. 
The field content supports the pullback action of the group of diffeomorphisms, 
\begin{equation}
\label{Diff-trsf-fields}
\begin{aligned}
    & \upphi^\psi := \psi^* \upphi, \quad
    \psi \in \Diff(M),
    \\[1mm]
    & \text{i.e.} \quad \lbrace{ A^\psi, \varphi^\psi, g^\psi \rbrace} := \lbrace{ \psi^* A, \psi^* \varphi,\psi^* g \rbrace}.
\end{aligned}
\end{equation}
A dressing field for diffeomorphisms is a smooth map
\begin{align}
\label{diffeo-dressing-field}
    \upsilon: N \rightarrow M, \qquad \text{such that} \quad \upsilon^\psi := \psi\- \circ \upsilon, 
\end{align}
for any $\psi \in \Diff(M)$.
A dressing field should be extracted from the field content of the theory, namely it should be a field-dependent dressing field $\upsilon = \upsilon[\upphi]$, so that $\upsilon^\psi := \upsilon (\psi^* \upphi) = \psi\- \circ \upsilon[\upphi]$, allowing a relational interpretation of the dressed variables.

Given such a dressing field $\upsilon$, the dressed fields are defined as
\begin{equation}
\label{diffeodressedfields}
\begin{aligned}
    & \upphi^\upsilon := \upsilon^* \upphi , \\[1mm]
    & \text{i.e.} \quad \lbrace{ A^\upsilon, \varphi^\upsilon, g^\upsilon \rbrace} = \lbrace{ \upsilon^* A, \upsilon^* \varphi,\upsilon^* g \rbrace},
\end{aligned}
\end{equation}
and are $\Diff(M)$-invariant by construction. 
Here we are using the DFM rule of thumb for the case of diffeomorphisms: 
To dress fields or functional thereof, we compute first their transformation under diffeomorphisms, then formally substitute $\psi \in \Diff(M)$ with the dressing field $\upsilon$. 
For $\upsilon=\upsilon[\upphi]$, the dressed fields \eqref{diffeodressedfields} are manifestly relational variables, i.e. 
they represent $\Diff(M)$-invariant relations among the physical spatio-temporal d.o.f. embedded in $\upphi$. 
Observe that they are not objects defined on the ``bare" manifold $M$.

\smallskip

In fact, the dressed fields \eqref{diffeodressedfields} live on  
field-dependent \emph{dressed regions}, defined by
\begin{align}
\label{fielddepdrreg}
    U^\upsilon = U^{\upsilon[\upphi]} 
    := \upsilon[\upphi]\- (U),
\end{align}
with $\upsilon\-$ the inverse map of $\upsilon$, such that $\upsilon \circ \upsilon\-=\id_M$. 
These are $\Diff(M)$-invariant:
Indeed, defining the action of $\Diff(M)$ on $U\subset M$ as $U \mapsto U^\psi:=\psi\- \circ U$, we find
\begin{equation}
\label{inv-dressed-regions}
\begin{aligned}
    (U^\upsilon)^\psi & = (\upsilon[\upphi]^\psi)\- \circ (U^\psi) \\
    & = \upsilon[\upphi]\- \circ \psi \circ \psi\- \circ (U) = U^\upsilon.
\end{aligned}
\end{equation}
The reason for \eqref{fielddepdrreg}, and the justification of the claim that dressed fields $\upphi^\upsilon$ live on field-dependent dressed regions, comes from integration theory, see \cite{JTF-Ravera2025bdyDFM} for details.
The $\Diff(M)$-invariant regions $U^{\upsilon[\upphi]}$ represent the physical regions of spatio-temporal events, a physical spatio-temporal event being a field-dependent $\Diff(M)$-invariant point $x^{\upsilon[\upphi]}:= \upsilon[\upphi]\-(x) \in U^\upsilon$.
We may then call $M^\upsilon := \text{Im}(\upsilon\-)$ the manifold of physical spatio-temporal events. It does not exist independently from the fields $\upphi$. 
Hence, the physical spacetime, i.e. the physical Lorentzian manifold, is $(M^\upsilon,g^\upsilon)$. 

\bigskip

Let us finally spend a few words about the dynamics.
The Lagrangian of the theory is $\Diff(M)$-covariant, and so are the field equations $E(\upphi)=0$ derived from it. 
Given a dressing field $\upsilon$, the dressed Lagrangian $L(\upphi^\upsilon):=\upsilon^*L(\upphi)$ is strictly $\Diff(M)$-invariant by construction. 
The field equations for the dressed fields, $ E(\upphi^u)=0$,
have the same functional expression as the ``bare" ones \cite{JTF-Ravera2024gRGFT}, but have a well-posed Cauchy problem.

\paragraph{Scalar coordinatization via dressing}

The framework described above encompasses diverse variants of the so-called ``scalar coordinatization" in general-relativistic physics. 
For instance, in approaches à la Brown–Kucha\v{r} \cite{Brown:1994py, Rovelli:2001my,Frankel2011}, if matter is described effectively as a fluid (gas, particles, dust, etc.), it provides scalars from which one gets the dressing field to obtain the manifestly diffeomorphism-invariant formulation. 
In this case, the dressed metric represents the invariant, \emph{relational} structure instantiated between the d.o.f. of the metric and of the effective matter field. 
This is the conceptual basis we start from in the present paper. 
It is also conceptually close to the proposals in \cite{Frob:2021mpb,DiFilippo:2023suv}, applied to black hole physics.
In the approach à la Kretschmann–Komar \cite{Komar:1958ymq, Bergmann:1972ud}, instead, a dressing field is extracted from the d.o.f. of the bare metric \footnote{This produces what we call an instance of self-dressing.}.

\section{Physical coordinatization corrections to rotational velocity}\label{Physical coordinatization corrections to rotational velocity}

We consider a theoretical setup whose field content is given by $\upphi=\{g, \upvarphi\}$, where $g$ represents the metric field and matter is described as a fluid/dust supplying scalar fields  $\upvarphi : U\subseteq M \rarrow N=\RR^4$, $\varphi=\varphi^a$, $a=1,2,3,4$. 
The $\Diff(M)$-transformations are
\begin{align}
  \upphi^\psi
  =
  \{\psi^*g , \, \psi^*\upvarphi \}
  =
  \{ \psi^*g, \,\upvarphi \circ \psi \}, \quad \psi \in \Diff(M).
\end{align}
We identify a $\Diff(M)$-dressing field as 
\begin{align}
\upsilon=\upsilon[\upvarphi]:=\upvarphi^{-1}: \RR^4 \rarrow M.
\end{align}
Indeed, it transforms under $\Diff(M)$ as a dressing field has to, $\upsilon^\psi=\upsilon[\vphi^\psi]=\psi\- \circ \upsilon[\upvarphi]$, see \eqref{diffeo-dressing-field}. 
Note that this allows to define dressed regions $U^\upsilon := \upsilon\- (U)$, such that $(U^\upsilon)^\psi = U^\upsilon$, yielding a $\Diff(M)$-invariant, relational definition of physical regions of events via (scalar) matter fields as a reference physical system \footnote{The fact that the matter distribution is self-dressed (that is, coordinatized with respect to itself) is expressed by $\upvarphi^{\upsilon}=\id_{U^\upsilon
}$, meaning that it is (invariantly) at rest in its own reference frame.}.

\subsection{Dressed metric components}

Most importantly for the cosmological application we aim to provide in this paper, the dressing procedure yields the $\Diff(M)$-invariant dressed metric $g^{\upsilon}$, which encodes the geometric properties of $M^{\upsilon}$ -- we refer the reader to \cite{JTF-Ravera2025bdyDFM} for the explicit proof of the $\Diff(M)$-invariance of the dressed metric. 

The dressed metric can be understood as the physical gravitational field as measured in the coordinate system supplied by the matter distribution $\upvarphi$. 
In abstract index notation, we have
\begin{align}
\label{dressed-metric-index}
\b g_{ab} := g^\upsilon_{ab} = {J(\upsilon)_a}^\mu \,  g_{\mu \nu} \, {J(\upsilon)^\nu}_b ,
\end{align}
the Jacobian of the ``physical coordinatization" being
\begin{align}
J(\upsilon)=J(\upvarphi\-)=J(\upvarphi)\-\!=\left( \tfrac{\partial \upvarphi^a}{\partial x^\mu} \right)\-\!.
\end{align}
We may now consider perturbation theory and write the ``dressed coordinates" as
\begin{align}
\label{xadressed}
    \b x^a := \varphi^a = \delta^a_\mu x^\mu + \chi^a(x)
\end{align}
perturbatively, in terms of the ``bare" ones, $x^\mu$, and of a small deformation $\chi^a=\chi^a(x)$, parametrizing the 
``infinitesimal" dressing. 
Considering $\chi^a$ small enough indeed, we will neglect higher-order terms in $\chi^a$, focusing on the linear, 1st order, level. 
One can also write the inverse relation $x^\mu = \delta^\mu_a (\varphi^a - \chi^a)$.
Using \eqref{dressed-metric-index} and \eqref{xadressed}, we can cast the dressed metric in the following form:
\begin{align}
    \label{drmetricpert}
    \b g_{ab} = g_{ab} - 2 g_{\mu(a} \partial_{b)} \chi^\mu ,
\end{align}
where $g_{ab}$ is the bare metric in components and $\partial$ denotes the ordinary partial derivative.

\smallskip

For simplicity, we shall now consider a galactic setup with spherical symmetry, which can be described in terms of the (bare) Schwarzschild metric, focusing on the equatorial plane (i.e., with fixed coordinate $\theta=\pi/2$) \footnote{We consider the metric signature to be $(-,+,+,+)$.},
\begin{equation}
\label{schwbare}
    ds^2 = - f dt^2 + f^{-1} dr^2 + r^2 d \phi^2 ,
\end{equation}
with $f=f(r):=1-2 GM/r$, $G$ being the gravitational constant and $M$ the central mass. We have the bare metric components
\begin{equation}
\begin{aligned}
    & g_{tt} = -f, \quad g_{rr} = f^{-1}, \quad g_{\phi \phi} = r^2 , \\
    & g^{tt} = -f^{-1}, \quad g^{rr} = f, \quad g^{\phi \phi} = r^{-2} ,
\end{aligned}
\end{equation}
while all off-diagonal components of the bare metric identically vanish. 
This provides the baseline geometry.

We dress the metric \eqref{schwbare} as in \eqref{dressed-metric-index}, taking the bare Schwarzschild metric components as $g_{\mu \nu}$. Note that dressed metric thus obtained has, in principle, also off-diagonal components. The functional expression of the dressed components, consisting of perturbative deformations of the bare ones, is different with respect to the original Schwarzschild one. In particular, setting, for simplicity and in compatibility with the assumed cosmological setting, $\partial_\phi \chi^a=0$ and $\partial_t \chi^\phi=0$, we get the relevant non-vanishing dressed components
\begin{equation}
\begin{aligned}
    \b g_{tt} & = g_{tt} - 2 g_{tt} \partial_t \chi^t = - f (1-2\partial_t \chi^t) , \\
    \b g_{\phi \phi} & = r^2, 
\end{aligned}
\end{equation}
while, e.g., under the above assumptions, $\b g_{t \phi}=0$.
Observe that the dressing operation thus yields an effective mass
\begin{align}
\label{Meff}
    M_{\text{eff}} = M (1-2\partial_t \chi^t) + \frac{r}{G} \partial_t \chi^t ,
\end{align}
where $\partial_t \chi^t$ is a small deformation (one can in fact assume $0 < \partial_t \chi^t < 1/2$), which, if constant, gives a contribution to the effective mass proportional to $r$. Notice also that, since, from \eqref{xadressed},
\begin{align}
\label{rbareintermsofdressed}
    r = \varphi^r - \chi^r = \b r - \chi^r ,
\end{align}
the effective mass can be rewritten in term of the dressed coordinate $\b r$, neglecting higher-order terms in the $\chi$ variables, as
\begin{align}
\label{Meffbr}
    M_{\text{eff}} \approx M (1-2\partial_t \chi^t) + \frac{\b r}{G} \partial_t \chi^t ,
\end{align}
which therefore has the same functional expression as \eqref{Meff} for the radial coordinate, respectively bare and dressed. 

Using \eqref{xadressed}, we also rewrite the dressed metric components in terms of the dressed coordinates. 
This because the dressed theory is the one to be compared with observations, and it shall be entirely expressed in terms of the dressed variables. We obtain 
\begin{equation}
\label{dressedcompmetricbr}
\begin{aligned}
     \b g_{tt} & = - \left[ 1 - \frac{2GM}{(\b r-\chi^r)} \right] \left(1 - 2 \partial_t \chi^t \right) , \\
     \b g_{\phi \phi} & = (\b r - \chi^r)^2 \approx \b r (1-2 \chi^r), 
\end{aligned}
\end{equation}
neglecting $\mathcal{O}^2(\chi^r)$ terms. From these one also derives
\begin{equation}
\label{drmetrcompderiv}
\begin{aligned}
    \frac{d \b g_{tt}}{d \b r} & = \frac{2 GM (-1+ \partial_t \chi^t)}{(\chi^r-\b r)^2} , \\
    \frac{d \b g_{\phi \phi}}{d \b r} & = 2 (\b r -\chi^r) ,
\end{aligned}
\end{equation}
which will be used in the following analysis of the dressed galaxy kinematics and test particles dynamics.

\subsection{Dressed rotational velocity}

Let us now consider the following (dressed) Lagrangian describing the dynamics of a test particle:
\begin{align}
    \bar{\mathcal{L}} = \frac{1}{2} \b g_{ab} \bar{\dot{x}}^a \bar{\dot{x}}^b ,
\end{align}
where $\bar{\dot{x}}^a$ denotes the dressed four-velocity. 
From it we can derive the dressed energy and angular momentum:
\begin{align}
    \b p_t = \frac{\partial \bar{\mathcal{L}}}{\partial \bar{\dot{t}}} = - \b E , \quad 
    \b p_\phi = \frac{\partial \bar{\mathcal{L}}}{\partial \bar{\dot{\phi}}} = \b L .
\end{align}
Thus, considering, for simplicity, circular orbits, $\bar{\dot r}=0$, which implies 
\begin{align}
\label{EVeff}
    \bar{E}^2=\b V_{\text{eff}}(\b r) ,
\end{align}
where $\b V_{\text{eff}}(\b r)$ is the dressed effective potential, together with $\b g_{t \phi}=0$, we get
\begin{align}
\label{bphidot-btdot}
    \bar{\dot{\phi}} = \frac{\b L}{\b g_{\phi\phi}} , \quad \bar{\dot{t}} = - \frac{\b E}{\b g_{tt}} .
\end{align}
We focus on the timelike case, for which we have
\begin{align}
    \b g_{\mu \nu} \bar{\dot{x}}^\mu \bar{\dot{x}}^\nu = -1 = \b g_{tt} \bar{\dot{t}}^2 + \b g_{\phi \phi} \bar{\dot{\phi}}^2 ,
\end{align}
and, using \eqref{EVeff} and \eqref{bphidot-btdot}, we find the following expression of the dressed effective potential:
\begin{align}
    \b V_{\text{eff}} (\b r) = - \b g_{tt} \left(1 + \frac{{\bar{L}}^2}{\b g_{\phi \phi}} \right) .
\end{align}
Then, solving the stability requirement $\frac{d \b V_{\text{eff}}}{d \b r}=0$, we get
\begin{align}
    {\bar{L}}^2 = \frac{\frac{d\b g_{tt}}{d \b r}\b g^2_{\phi \phi}}{\frac{d\b g_{\phi \phi}}{d \b r}\b g_{t t}- \frac{d\b g_{tt}}{d \b r}\b g_{\phi \phi}} .
\end{align}
We can express the \emph{dressed rotational velocity} as $\b v= \b r \bar{\dot{\phi}}= \b r \, \frac{\b L}{\b g_{\phi \phi}}$, that therefore is
\begin{align}
\label{drrotvel}
    \b v^2 = \left( \b r \, \frac{\b L}{\b g_{\phi \phi}} \right)^2 = \frac{\b r^2 \, \frac{d\b g_{tt}}{d \b r}}{\frac{d\b g_{\phi \phi}}{d \b r}\b g_{t t}- \frac{d\b g_{tt}}{d \b r}\b g_{\phi \phi}} .
\end{align}
Finally, we substitute in \eqref{drrotvel} eqs.
\eqref{dressedcompmetricbr} and \eqref{drmetrcompderiv} for the dressed metric components, obtaining
\begin{align}
\label{drrotvelchi}
    \b v^2 = \frac{GM \b r}{6 GM \chi^r -3 (\chi^r+GM)\b r+\b r^2} .
\end{align}
where, again, we have neglected $\mathcal{O}^2(\chi^r)$ terms.
Then, in the weak-field approximation ($\b r >> 2GM$, $1-\frac{2GM}{\b r} \approx 1$), \eqref{drrotvelchi} boils down to
\begin{align}
\label{drrotvelchiwf}
    \b v^2 \approx \frac{GM}{\b r - 3 \chi^r} \approx \frac{GM}{\b r} + \frac{3GM \chi^r}{\b r^2} .
\end{align}
We observe that the first term in \eqref{drrotvelchiwf} reproduces the Keplerian behavior, while the second term is a correction due to the presence of $\chi^r$. In particular, if $\chi^r \propto \b r^2$, the second term is positive and constant. We are thus left with
\begin{align}
\label{drrotvelfin}
    \b v^2 = \frac{GM}{\b r} + \b v^2_0 , \quad \b v_0 = \text{ constant} ,
\end{align}
where $\b v_0^2 = 3 GM k$, $k>0$ being the proportionality constant in 
\begin{align}
\label{chircond}
    \chi^r = k \b r^2 .
\end{align}
The expression in \eqref{drrotvelfin} for the dressed rotational velocity thus contains the (dressed version of the) usual Keplerian term plus a corrective term induced by the dressed formulation which effectively emulates the contribution expected from DM, raising the galaxy rotation curve. 

Since $\chi^r=\chi^r(r)$, the condition on $\chi^r$ imposed to obtain this result, \eqref{chircond}, shall be rewritten, using \eqref{rbareintermsofdressed} and neglecting $\mathcal{O}^2(\chi^r)$ terms, as
\begin{align}
\label{chicondr}
    \chi^r = \frac{k r^2}{1-2k r}.
\end{align}
As a final step of our analysis, we check the consistence of the conditions implemented on $\chi^a$, giving the scalar profile, and its derivatives with the (dressed) four-velocity of a cosmological fluid or dust field.

\section{Scalar profile and four-velocity of dust field}\label{Scalar profile and four-velocity of dust field}

We define the effective dressed four-velocity of the (co-moving) dust field as
\begin{align}
    \b u^a := \frac{d\varphi^a}{d \tau} ,
\end{align}
such that, in terms of the bare four-velocity $u^\mu := \frac{dx^\mu}{d\tau}=(1,0,0,0)$, $u^\mu u_\mu = -1$, $\tau$ being the proper time, we write
\begin{align}
    \b u^a := \frac{d\varphi^a}{d \tau} = \frac{\partial \varphi^a}{\partial x^\mu} \frac{d x^\mu}{d \tau} = \frac{\partial \varphi^a}{\partial x^\mu} u^\mu .
\end{align}
Recall that we are considering motion in the equatorial plane, $\theta=\pi/2$ and $d \theta/d\tau=0$, and circular orbits.
Then, using the fact that $\varphi^a = \delta^a_\mu x^\mu + \chi^a$, we get
\begin{align}
\label{dr4vel}
    \b u^a = \left( \delta^a_\mu + \frac{\partial \chi^a}{\partial x^\mu} \right) u^\mu .
\end{align}
The conditions we have obtained on $\chi^a$ and its derivatives to get a constant corrective term in the rotational velocity squared, raising the galaxy rotation curve, can be collected as follows, together with their relevant implications:
\begin{equation}
\label{condschi}
\begin{aligned}
    & \partial_\phi \chi^a = 0 , \quad \partial_t \chi^\phi = 0 ; \\
    & \partial_t \chi^t = \epsilon, \quad \text{$\epsilon$ small constant} \quad \Rightarrow \quad \chi^t = \epsilon t + F, \\
    & \chi^r = \frac{k r^2}{1-2k r} \quad \Rightarrow \quad \partial_t \chi^r =  \partial_\phi \chi^r = \partial_\theta \chi^r = 0 , \\
    & \,\phantom{\chi^r = \frac{k r^2}{1-2k r} \quad \Rightarrow \quad} \partial_r \chi^r = \frac{2 k r}{(1-2kr)^3} ,
\end{aligned}
\end{equation}
where, in principle, $F=F(r,\phi, \theta)$ (one can also consider, for simplicity, $F$ to be a function of $r$ only). One can now see that eqs. \eqref{condschi} are consistent with the description of the dressed four-velocity of the dust field.
In particular, for consistency with the dust four-velocity description in the bare case, one would expect $\b u^a=(\b u^t,0,0,0)$. Which is indeed the case, as we can see from \eqref{dr4vel}, using \eqref{condschi}:
\begin{align}
    \b u^a = (1+\epsilon)u^\mu = \left( (1+\epsilon)u^t,0,0,0 \right) = (1+\epsilon,0,0,0) ,
\end{align}
with $\epsilon$ representing the small, constant deformation coming from the dressing operation \footnote{One can then also show that, in the weak field limit, $\b u^a \b u_a\approx -1$.}. 

\begin{figure*}[ht]
    \centering
    \begin{subfigure}{0.49\textwidth}
        \centering
        \includegraphics[width=\textwidth]{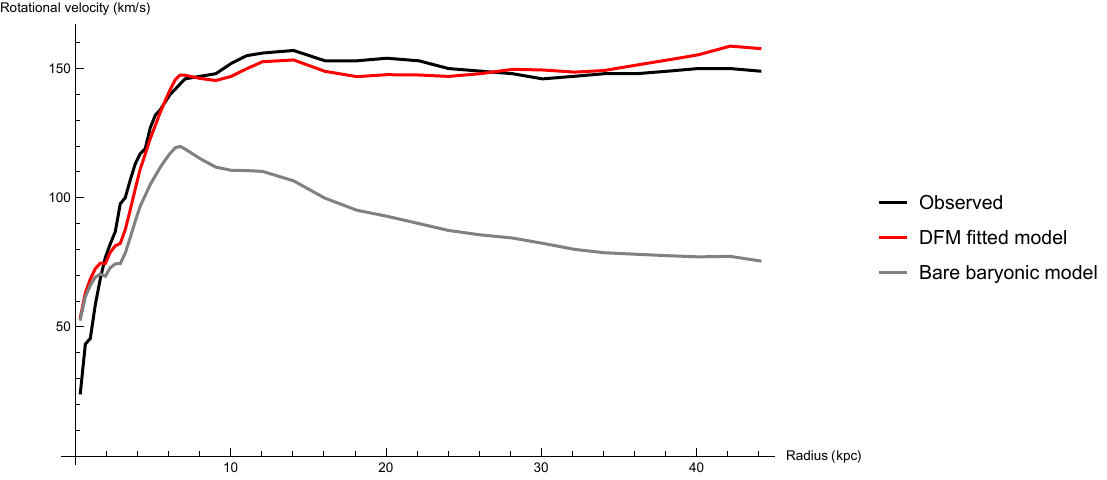}
        \caption{NGC 3198 rotation curves.}
        \label{fig:plotFitNGC3198}
    \end{subfigure}
    \hfill 
    \begin{subfigure}{0.49\textwidth}
        \centering
        \includegraphics[width=\textwidth]{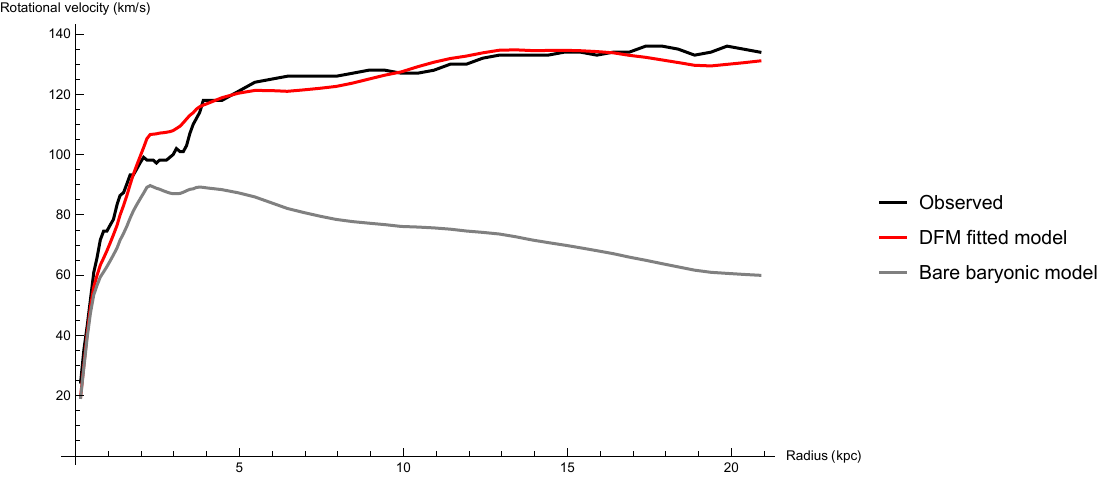}
        \caption{NGC 2403 rotation curves.}
        \label{fig:plotFitNGC2403}
    \end{subfigure}
    \caption{Rotation curves of NGC 3198 (a) and NGC 2403 (b). 
    The black lines show observed velocities from the SPARC database \cite{Lelli:2016zqa}. The red lines represent the DFM-derived rotation curve. The gray lines show the bare baryonic model prediction, assuming a Keplerian law.}
    \label{fig:plotsFitNGC3198andNGC2403}
\end{figure*}

\section{Phenomenological comparison of DFM to observed rotation curves}\label{Phenomenological comparison of DFM to observed rotation curves}

To assess the phenomenological viability of the rotation curve we have derived through the DFM, which can be read from \eqref{drrotvelfin}, we may compare it to observed rotation curves of spiral galaxies.

We select NGC 3198 and NGC 2403 -- the latter is less massive than the former -- from the SPARC database \cite{Lelli:2016zqa}, as both are nearby late-type spirals with extended, high-quality HI rotation curves and well-constrained inclinations. Their kinematic regularity and extensive use as benchmark systems in studies of disk dynamics make them well suited for testing the phenomenological viability of our derived rotation curves.
In the Supplemental Material \ref{Supplemental Material: Further galaxy plots and fits of DFM velocity profiles} we show the plots of the observed rotation curves of NGC 3198 and NGC 2403.

From the SPARC database we take the rotational velocities of the gas, disk, and bulge components, derived from the surface-brightness distribution, and use them to estimate the corresponding baryonic mass contributions. 
These are then rescaled by the mass–to–light corrective factors $\Upsilon_{\text{gas}} = 1.33 \, M_\odot/L_\odot$, $\Upsilon_{\text{disk}} = 0.70 \, M_\odot/L_\odot$, and $\Upsilon_{\text{bulge}} = 0.70 \, M_\odot/L_\odot$ (see \cite{Walter2008} and \cite{deBlok:2008wp}), yielding the physical baryonic mass of each component. 
Summing these we get the total baryonic mass of the galaxies.
In the Supplemental Material \ref{Supplemental Material: Further galaxy plots and fits of DFM velocity profiles} we also provide the baryonic mass profile plots of NGC 3198 and NGC 2403.

The DFM model (dressed) rotational velocity, to be compared with observations, is given by 
\begin{align}
\label{barvDFMmodel}
    \bar{v} = \sqrt{\frac{GM}{\bar{r}} + \bar{v}_0^2},
\end{align}
where $G = 4.302 \times 10^{-6} \, \text{kpc} \, (\text{km/s})^2 \, M_\odot^{-1}$ in cosmological units. 

In Fig. \ref{fig:plotsFitNGC3198andNGC2403} we present the fit of the DFM velocity profile to the observed rotation curves, using as input mass $M$ the baryonic mass previously derived. 
For completeness, and to highlight the departure from the purely Newtonian expectation, we also plot the Keplerian velocity profile corresponding to the bare baryonic model (for the same baryonic mass $M$).
The red lines represent the DFM-derived rotation curve with best-fit parameter $k = 0.0254 \, \text{kpc}^{-1}$ for NGC 3198 and $k = 0.0604 \, \text{kpc}^{-1}$ for NGC 2403, respectively, in both cases with a precision of $10^{-5}$.
The quality of the fits is good, with coefficients of determination $R^2 = 0.995609$ for NGC 3198 and $R^2 = 0.99794$ for NGC 2403 \footnote{The coefficient of determination $R^2$ measures the fraction of the total variance in the observed data that is captured by the model, with $R^2 = 1$ indicating a perfect fit and $R^2=0$ meaning the model explains no variance beyond the mean. We quantify the fit quality using the coefficient of determination $R^2$ rather than a reduced $\chi^2$, because the $\chi^2$ statistic assumes that the observational uncertainties are statistically reliable and Gaussian. In the SPARC data, the error bars vary in precision and are particularly large at small radii for NGC 3198 and at large radii for NGC 2403, violating these assumptions. The $R^2$ metric provides a more appropriate measure for evaluating the overall agreement in this context.}.

In the Supplemental Material \ref{Supplemental Material: Further galaxy plots and fits of DFM velocity profiles}, we also provide a comparison of the residuals (namely the difference between the values of the observed velocity and those provided by the DFM) between the dressed model and the bare baryonic Keplerian model. The DFM residuals are consistently closer to zero across most radii, indicating a significantly better agreement with the observed rotation curves than, e.g., the bare baryonic model.

This analysis demonstrates that the DFM framework can replicate the flat rotation curves observed in spiral galaxies at large radii -- with velocities stabilizing at values consistent with typical spiral galaxy observations \cite{Sofue:2000jx,Walter2008,deBlok:2008wp,Lelli:2016zqa} -- through a constant corrective term, without invoking DM in this context.

\section{Conclusions}
\label{Conclusions}

We have demonstrated that the DFM provides a relational, diffeomorphism-invariant framework that, when applied to galaxy dynamics, yields rotational velocities comprising a Keplerian term and a constant corrective term. 
This approach produces flat galaxy rotation curves, mirroring the effects typically attributed to DM without requiring additional unseen mass in this context. 
By constructing physical d.o.f. through DFM, our framework offers a novel perspective on the rotation curve anomaly, highlighting the potential of gauge-invariant formulations to capture observed galactic dynamics. 
While our analysis is limited to rotation curves and does not address broader DM phenomenology, such as the (baryonic) Tully-Fisher relation (BTFR), CMB fluctuations or gravitational lensing, these findings pave the way for further applications of DFM in cosmology, including cosmological perturbation theory and other observational tests, as planned in future work.

We have considered the Schwarzschild metric to provide
the baseline geometry, and 
matter described effectively as a cosmological fluid or dust, providing scalars from which we extracted the dressing field to obtain the manifestly
diffeomorphism-invariant formulation. The scalars can be thus seen as physical reference frame, whose profile, subject to conditions compatible with the dressed four-velocity of the dust field, allows for a corrective term
that raises the galaxy rotation curve.

To assess the phenomenological viability of the rotation curve we have derived through the DFM, we have also provided a phenomenological analysis, comparing our DFM-derived rotation curves to observed data for the spiral galaxies NGC 3198 and NGC 2403, using data from the SPARC database \cite{Lelli:2016zqa}. 
We have thus shown that the DFM model reproduces well the flat rotation curve at large radii, performing a one-parameter (the free parameter $k$ left of the DFM theoretical analysis) fit to the observed rotational velocities of NGC 3198 and NGC 2403.

Our analysis shows that the DFM model, using the purely baryonic mass as input, provides a good fit to the observed rotation curves of both galaxies, with best-fit parameters 
$k = 0.0254 \pm 10^{-5} \, \text{kpc}^{-1}$ for NGC 3198 and $k = 0.0604 \pm 10^{-5} \, \text{kpc}^{-1}$ for NGC 2403. 
The model reproduces the characteristic flattening of the curves at large radii, with velocities stabilizing around 150 km/s for NGC 3198 and approximately 135-140 km/s for NGC 2403, consistent with typical spiral galaxy observations. 
Residuals analysis (see \cite{SM}) further confirms that the DFM predictions closely match the observed velocities, outperforming the bare baryonic Keplerian model. 

We have examined several other cases and consistently found agreement with the results of the previous sections, as displayed in the Supplemental Material \ref{Supplemental Material: Further galaxy plots and fits of DFM velocity profiles}. These results support the phenomenological viability of the DFM framework in describing spiral galaxy rotation curves without invoking DM.

In forthcoming papers, we will further apply the DFM, encompassing the notion of coordinatisation through coupling to matter, to develop a manifestly relational, thus automatically $\Diff(M)$-invariant, version of cosmological perturbation theory \cite{Giesel:2007wi,Chiaffrino:2020akd,Frob:2023awn}, and to study other possible cosmological effects driven by dressing.

\begin{acknowledgments}
\vspace{-1mm}
J.F. is supported by the Austrian Science Fund (FWF), grant \mbox{[P 36542]}, 
and by the Czech Science Foundation (GAČR), grant GA24-10887S.
L.R. is supported by the 
GrIFOS research project, funded by the Ministry of University and Research (MUR, Ministero dell'Università e della Ricerca, Italy), PNRR Young Researchers funding program, MSCA Seal of Excellence (SoE), 
CUP E13C24003600006, ID SOE2024$\_$0000103.
\end{acknowledgments}



\section{Supplemental Material: Further galaxy plots and fits of DFM velocity profiles}\label{Supplemental Material: Further galaxy plots and fits of DFM velocity profiles}

This Supplemental Material provides plots of the observed rotational velocities of the galaxies NGC 3198 and NGC 2403, those of their baryonic mass profile, the analysis of the residuals for the dressed model fits, and additional fits of galaxy velocity profiles derived via the Dressing Field Method (DFM).

\subsection{Content of the Supplemental Material}\label{Content}

In Section \ref{Supplementary plots for the galaxies NGC 3198 and NGC 2403}, we provide the plots of the observed rotational velocities of the galaxies NGC 3198 and NGC 2403, together with their baryonic mass profile.
In this same section, we also include the the analysis of the residuals for the dressed model fits.
Section \ref{Further fits of DFM velocity profiles} contains 
the fits of the Dressing Field Method (DFM) velocity profiles to the observed rotation curves of other high-resolution galaxies from the SPARC database \cite{Lelli:2016zqa}, which further support the results presented in our paper.

\begin{figure*}[ht]
    \centering
    \begin{subfigure}{0.49\textwidth}
        \centering
        \includegraphics[width=\textwidth]{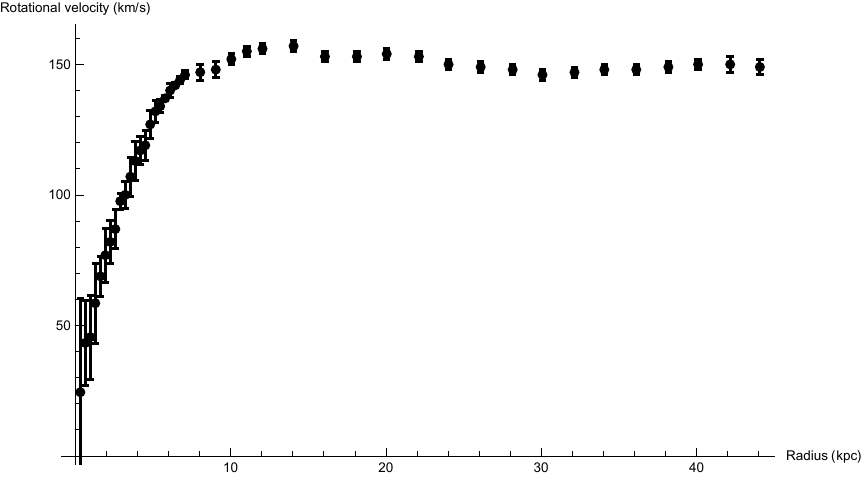}
        \caption{NGC 3198 observed rotational velocity.}
        \label{fig:plotVobsNGC3198}
    \end{subfigure}
    \hfill 
    \begin{subfigure}{0.49\textwidth}
        \centering
        \includegraphics[width=\textwidth]{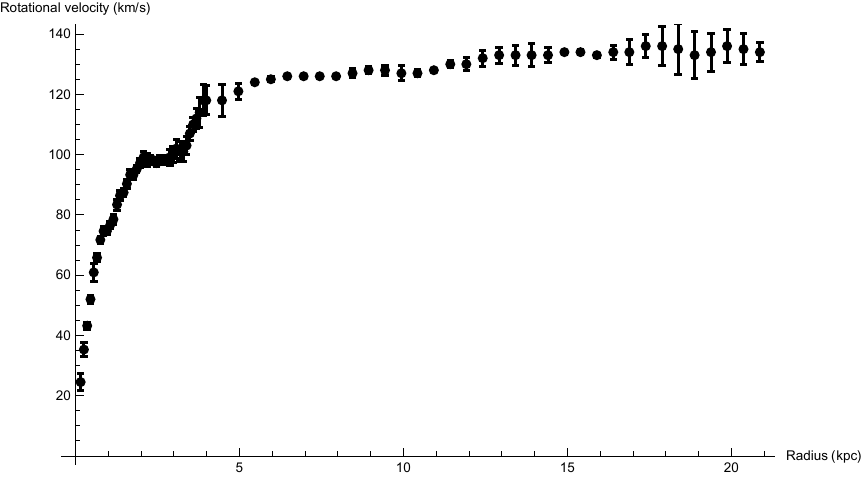}
        \caption{NGC 2403 observed rotational velocity.}
        \label{fig:plotVobsNGC2403}
    \end{subfigure}
    \caption{Observed rotational velocities of the galaxies NGC 3198 (a) and NGC 2403 (b). Data extracted from the SPARC database.}
    \label{fig:plotsVobsNGC3198andNGC2403}
\end{figure*}

\begin{figure*}[ht]
    \centering
    \begin{subfigure}{0.49\textwidth}
        \centering
        \includegraphics[width=\textwidth]{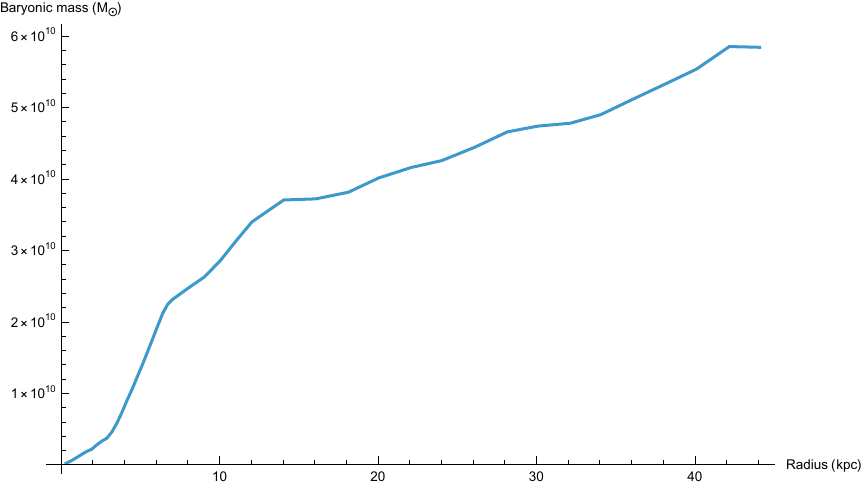}
        \caption{NGC 3198 baryonic mass profile.}
        \label{fig:plotMbarNGC3198}
    \end{subfigure}
    \hfill 
    \begin{subfigure}{0.49\textwidth}
        \centering
        \includegraphics[width=\textwidth]{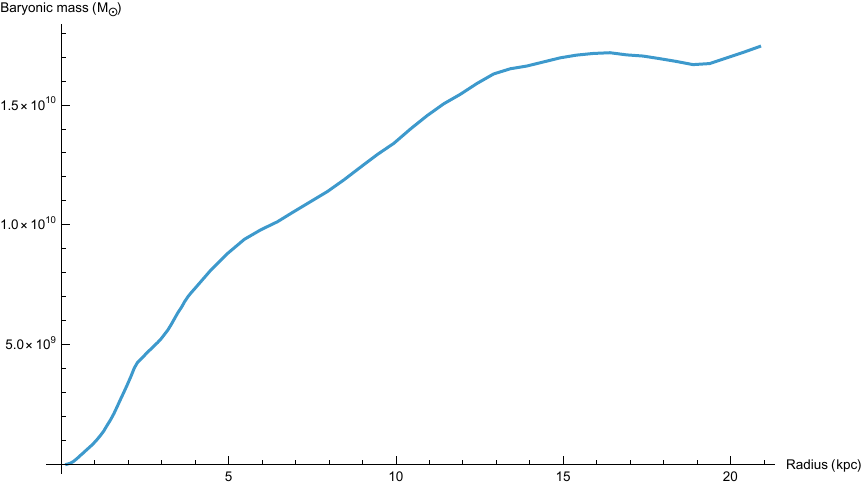}
        \caption{NGC 2403 baryonic mass profile.}
        \label{fig:plotMbarNGC2403}
    \end{subfigure}
    \caption{Baryonic mass profile of the galaxies NGC 3198 (a) and NGC 2403 (b), derived by summing gas, disk, and bulge mass contributions.}
    \label{fig:plotsMbarNGC3198andNGC2403}
\end{figure*}

\subsection{Supplementary plots for the galaxies NGC 3198 and NGC 2403}\label{Supplementary plots for the galaxies NGC 3198 and NGC 2403}

In Fig. \ref{fig:plotsVobsNGC3198andNGC2403} we provide the observed rotation curves of the galaxies NGC 3198 and NGC 2403, where we have plotted the rotational velocity as a function of radius, with data and error bars as provided in the SPARC database. 
Note that for NGC 3198 the uncertainties are relatively large at small radii, while for NGC 2403 the error bars increase significantly where the curve begins to flatten and at the outermost radii.
As expected for late–type spirals, both galaxies exhibit the characteristic flattening of the rotation curve at large radii, which serves as the standard phenomenological feature to be reproduced by any viable dynamical model.

As reported in the main paper, we have then estimated the baryonic mass contributions from gas, disk, and bulge.
These have been then rescaled by the mass–to–light corrective factors $\Upsilon_{\text{gas}} = 1.33 \, M_\odot/L_\odot$ (see the THINGS survey \cite{Walter2008}), $\Upsilon_{\text{disk}} = 0.70 \, M_\odot/L_\odot$, and $\Upsilon_{\text{bulge}} = 0.70 \, M_\odot/L_\odot$ (cf. \cite{Walter2008} and \cite{deBlok:2008wp}), yielding the physical baryonic mass of each component. 

Note that, for the gas, adopting $\Upsilon_{\text{gas}} = 1.00$ would omit the standard helium (and metals) correction and underestimate the baryonic gas mass. For the stellar disk, $\Upsilon_{\text{disk}} = 0.70$ at $3.6 \, \mu\text{m}$ lies within the range found in precise rotation–curve fits (e.g. \cite{deBlok:2008wp}) and is consistent with stellar–population synthesis models for intermediate–age disks, while remaining below the ``maximum disk'' limit ($\sim 1.00$) inferred in baryonic–only or MOND fits. For the bulge, we follow the THINGS survey.

Summing these contributions, we have derived the total baryonic mass of the galaxies.
Fig. \ref{fig:plotsMbarNGC3198andNGC2403} shows the baryonic mass profiles of NGC 3198 and NGC 2403 we obtained, plotted with respect to galactocentric radius.

\subsubsection{Dressed model residuals}

\begin{figure*}[ht]
    \centering
    \begin{subfigure}{0.49\textwidth}
        \centering
        \includegraphics[width=\textwidth]{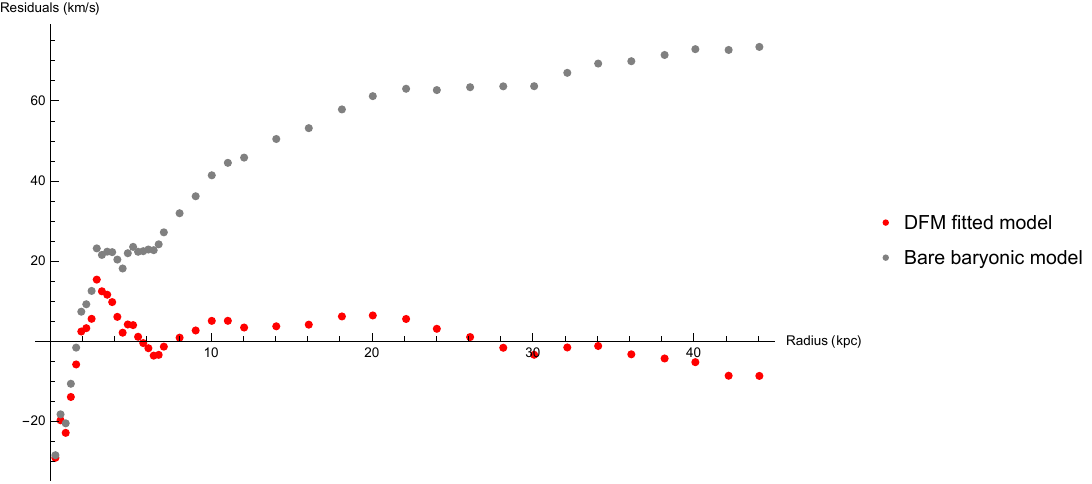}
        \caption{Comparison of residuals for NGC 3198.}
        \label{fig:plotResNGC3198}
    \end{subfigure}
    \hfill 
    \begin{subfigure}{0.49\textwidth}
        \centering
        \includegraphics[width=\textwidth]{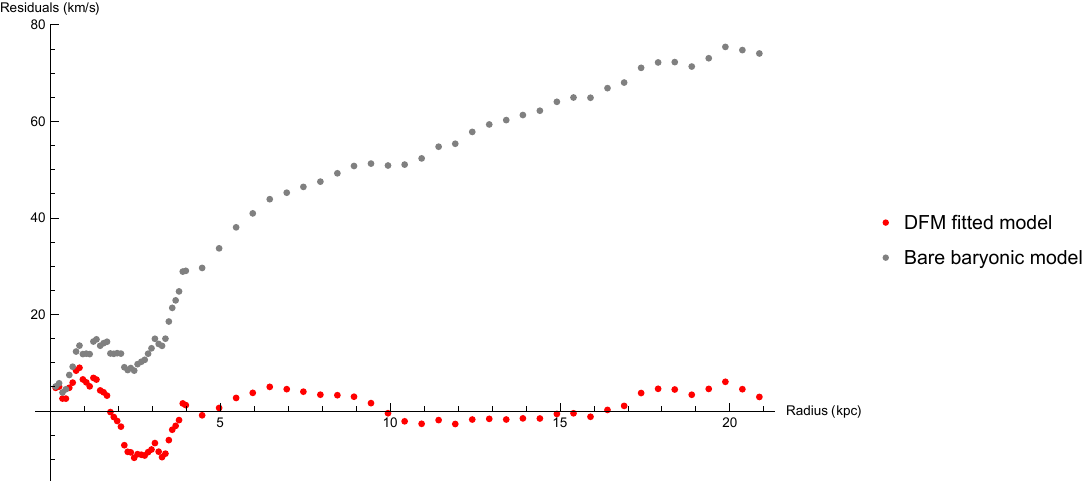}
        \caption{Comparison of residuals for NGC 2403.}
        \label{fig:plotResNGC2403}
    \end{subfigure}
    \caption{Comparison between the dressed model residuals (red points) and those of the bare baryonic model (gray points), for the galaxies NGC 3198 (a) and NGC 2403 (b).}
    \label{fig:plotsResNGC3198andNGC2403}
\end{figure*}

In Fig. \ref{fig:plotsResNGC3198andNGC2403} we provide a comparison of the residuals between the dressed model and the bare baryonic Keplerian model, for both NGC 3198 and NGC 2403. 
Despite the data quality being variable across regions, it is evident that the DFM residuals are consistently closer to zero across most radii, indicating a significantly better agreement with the observed rotation curves than, e.g., the bare baryonic model.
This analysis complements our study concerning the fits of the DFM velocity profiles of the galaxies NGC 3198 and NGC 2403.

\subsection{Further fits of DFM velocity profiles}\label{Further fits of DFM velocity profiles}

In this section, we provide additional fits of the DFM velocity profiles to high-resolution galaxy rotation curves from the SPARC database. 
This supplementary analysis is restricted to rotation curves containing at least 20 data points, thereby ensuring statistically more robust results, and with error bars consistent with high-quality datasets.

\begin{figure}[ht]
    \centering
    \includegraphics[width=0.92\textwidth]{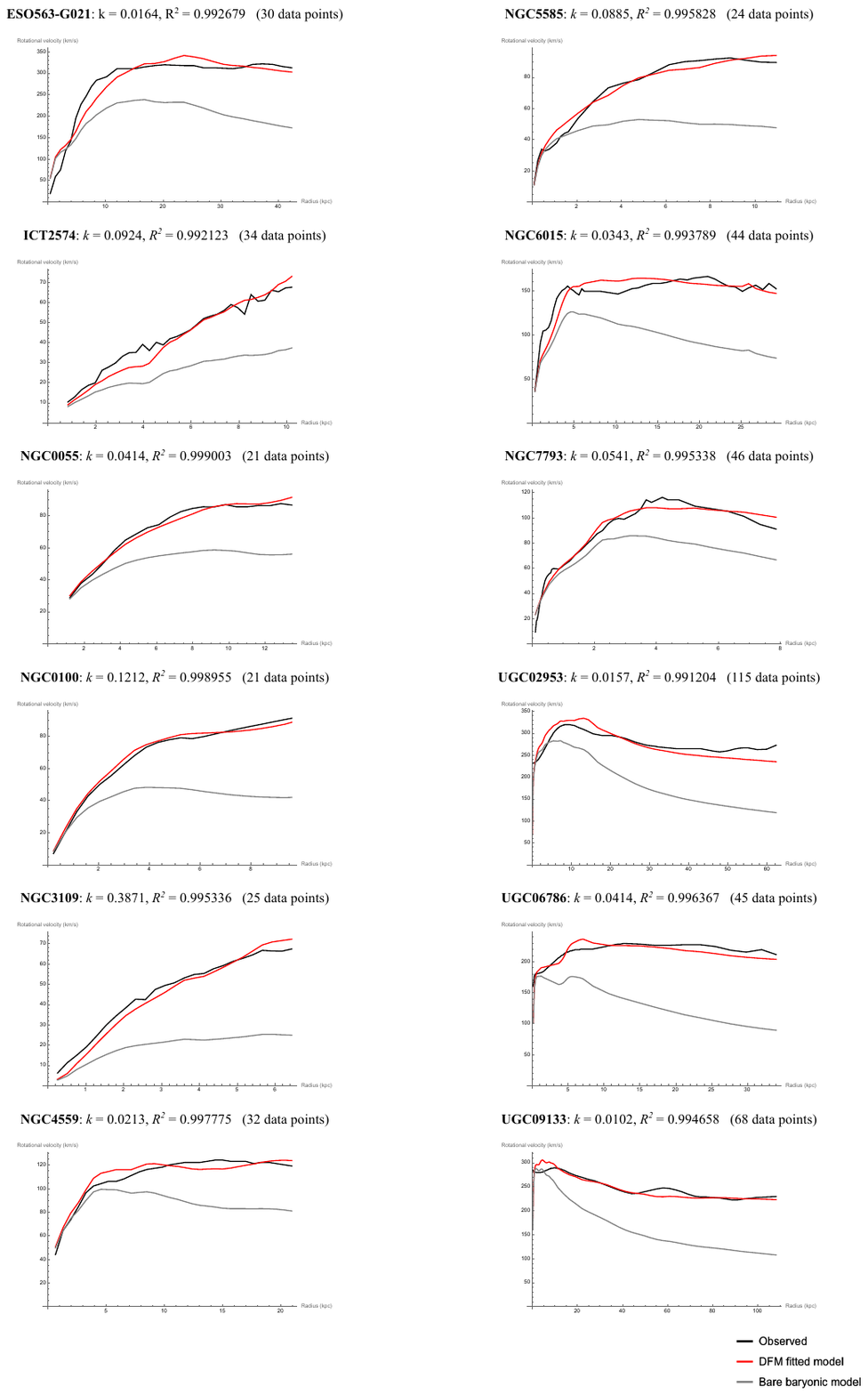}
    \caption{Other high-resolution rotation curves analyzed.}
    \label{fig:other_galaxies}
\end{figure}

Fig. \ref{fig:other_galaxies} presents a representative sample of galaxies analyzed, where we list the galaxy name, with number of available data points in the SPARC database, the fitted parameter $k$, and the $R^2$ (also reported in Table \ref{tab:kR2}), confirming that the DFM provides consistently good fits for rotation curves constructed from high-quality data.

\begin{table}[ht]
\caption{Fitted parameter $k$ and $R^2$ for the fits of the DFM velocity profiles of the high-resolution galaxies displayed in Fig. \ref{fig:other_galaxies}.}
\label{tab:kR2}
\begin{ruledtabular}
\begin{tabular}{lccc}
Galaxy Name & N. of Data Points & $k$ & $R^2$ \\
\hline
ESO563-G021 & 30 & 0.0164 & 0.992679 \\
ICT2574 & 34 & 0.0924 & 0.992123 \\
NGC0055 & 21 & 0.0414 & 0.999003 \\
NGC0100 & 21 & 0.1212 & 0.998955 \\
NGC3109 & 25 & 0.3871 & 0.995336 \\
NGC4559 & 32 & 0.0213 & 0.997775 \\
NGC5585 & 24 & 0.0885 & 0.995828 \\
NGC6015 & 44 & 0.0343 & 0.993789 \\
NGC7793 & 46 & 0.0541 & 0.995338 \\
UGC02953 & 115 & 0.0157 & 0.991204 \\
UGC06786 & 45 & 0.0414 & 0.996367 \\
UGC09133 & 68 & 0.0102 & 0.994658 \\
\end{tabular}
\end{ruledtabular}
\end{table}

Let us conclude by mentioning that we have also examined rotation curves from the SPARC database with lower-quality data (meaning fewer data points with bigger systematic errors). 
In those cases,
the DFM fits remain reasonably good, with no significant discrepancies compared to the results we presented in the main article and in this section,
though, predictably, they are not as accurate as those for high-quality data.


\clearpage

\bibliography{bibdmdfm.bib}

\end{document}